\documentclass{article}
\usepackage[twocolumn]{emulateapj}
\submitted{To appear in ApJ}

\lefthead{\sc Blakeslee \& Metzger}
\righthead{\sc Arc in Abell 2124}
\def\lta{\mathrel{\rlap{\lower 3pt\hbox{$\mathchar"218$}}
     \raise 2.0pt\hbox{$\mathchar"13C$}}}
\def\gta{\mathrel{\rlap{\lower 3pt\hbox{$\mathchar"218$}}
     \raise 2.0pt\hbox{$\mathchar"13E$}}}
\def\kms{km~s$^{-1}$}

\def\hkpc{$h^{-1}\,$kpc}
\def\etal{{\it et~al.}} 
\def\pf{\ifmmode{{\hbox{\sc psf}}}\else{{\sc psf}}\fi}
\def\mM{\ifmmode(m{-}M)\else$(m{-}M)$\fi}
\def\msun{\ifmmode{\hbox{M$_\odot$}}\else{M$_\odot$}\fi}

\begin{document}
\tighten

\title{A Lensed Arc in the Low Redshift Cluster Abell~2124\altaffilmark{1}}

\author{John P. Blakeslee and Mark R. Metzger}
\affil{Palomar Observatory, California Institute of Technology,
Mail Stop 105-24, Pasadena,~CA~91125}
\authoremail{jpb@astro.caltech.edu}
\authoremail{mrm@caltech.edu}
 
\altaffiltext{1}{Based on observations obtained at the 
W.M. Keck Observatory, operated as a scientific partnership
by the California Institute of Technology, the University of California,
and the National Aeronautics and Space Administration.
}

\begin{abstract}
We report the discovery of an arc-like object
27\arcsec\ from the center of the cD galaxy in the 
redshift $z=0.066$ cluster A2124.
Observations with the Keck~II telescope reveal that the object is a
background galaxy at $z=0.573$, apparently
lensed into an arc of length $\sim\,$8\farcs5
and total R~magnitude $m_R = 20.86\pm0.07$.
The width of the arc is resolved; we estimate it to be 
$\sim\,$0\farcs6 after correcting for seeing. 
A lens model of the A2124 core mass distribution consistent with the
cluster galaxy velocity dispersion reproduces the observed 
arc geometry and indicates a magnification factor $\gta\,$9.
With this magnification, the strength of the 
[OII] $\lambda$3727 line implies a star-formation rate of
$\hbox{SFR}\sim0.4\;h^{-2}\,\msun\,$yr$^{-1}$.
A2124 thus appears to be the lowest redshift cluster known to exhibit
strong lensing of a distant background galaxy.

\end{abstract}


\section{Introduction}

The vast majority of known gravitational arcs have been identified
in galaxy clusters at redshifts $z>0.2$ (see for instance the
tabulations by LeF\`evre \etal\ 1994 and Kneib \& Soucail 1996).
They have been used to place constraints on the mass distributions
within these clusters free from any assumptions regarding hydrostatic
or dynamical equilibrium (Narayan \& Bartelmann 1998 give a recent
review).  Such instances of strong lensing also provide the opportunity
to study the properties of the distant background galaxies that are
magnified by the foreground cluster lenses.

One cluster with $z<0.1$ proposed to exhibit a gravitational arc is
A3408 at $z=0.042$ (Campusano \& Hardy 1996).  Campusano, Kneib, \&
Hardy (1998) further investigated the possibility that the spiral
galaxy at $z=0.073$ located 50\arcsec\ from the central elliptical in
A3408 is being lensed.  They found that the lensing must be fairly
weak, with a magnification factor $\lta\,$1.7.  Even so, lensing in
low-redshift clusters in general provides the opportunity to study
core mass distributions at higher spatial resolutions than that
afforded by the moderate to high-redshift clusters that have been the
focus of most lensing studies so far.

Abell~2124 is a richness class~1 cluster at a redshift $z = 0.066$.
Hill \& Oegerle (1993) measured velocities for 66 galaxies 
with $17,800<cz<23,700$ \kms\ in the field of A2124 and reported a
cluster dispersion of $\sigma_{cl}=1180$~\kms; they noted that the high 
dispersion was primarily due to four galaxies with $cz\sim23,500$ \kms.
Fadda \etal\ (1996) reanalyzed the velocity data
and concluded $\sigma_{cl}=878^{+90}_{-72}$~\kms\ from 61 cluster members.
This latter value is more appropriate to A2124's fairly modest X-ray
luminosity $L_X = 1.4\times10^{44}$ ergs~s$^{-1}$
in the 0.1--2.4~keV band (Ebeling \etal\ 1996).
However, the cluster velocity dispersion profile shown by Fadda \etal\
appears to rise to $\sigma_{cl}\gta1100$ \kms\ within 
$r\lta100$ \hkpc\ (where $h$ is the Hubble constant in units of
100\,\kms~Mpc$^{-1}$).

As a Bautz-Morgan type~I, Rood-Sastry class cD cluster
(Abell \etal\ 1989; Struble \& Rood 1987), A2124 is dominated
by a single giant cD galaxy, UGC~10012.  The cD has a central
stellar velocity dispersion of $305\pm15$ \kms\ 
and a ``secondary nucleus'' at a projected separation
of 11.2~\hkpc\ with a relative
velocity of $949\pm14$~\kms\ (\cite{BT92}). 
The cD itself has no significant peculiar velocity with 
respect to the cluster mean (\cite{OH94}).  

\section{Observations}

We observed the A2124 cD as part of a project to extend
to larger distances and denser environments a previous survey 
of the globular cluster populations of brightest cluster galaxies
(Blakeslee, Tonry, \& Metzger 1997).
In the course of inspecting the final image, we noticed a narrow
arc-like object 27\arcsec\ along the major axis from the cD center.
The orientation of the arc suggested that it might be lensed, so
we carried out further observations to test this hypothesis.

\subsection{Imaging Data}

\begin{figure*}[t]\epsscale{0.85}
\plotone{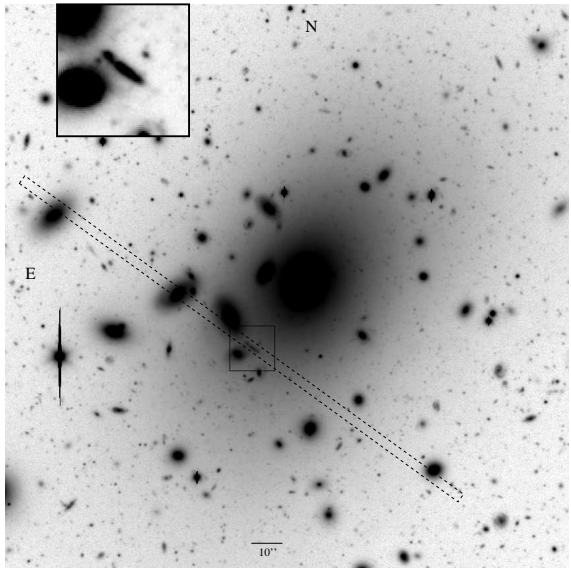}\caption{ \small
The central $3{\,\times\,}3$ arcminutes of the
$8400\,$s LRIS $R$-band image centered near the A2124 cD
is displayed with a logarithmic stretch.
The dotted line shows schematically the placement of the
LRIS slit for the spectroscopic observations presented here.
A $14\arcsec\times14\arcsec$ box drawn around the arc is
enlarged by a factor of three in the inset.  The inset uses
a linear stretch and has had a smooth background fit to the
cD halo subtracted.
\label{img}}
\end{figure*}

The cD galaxy in A2124 was imaged in the $R$~band on the night of
28~April 1997 with the Low Resolution Imaging Spectrograph (LRIS, Oke
\etal\ 1995) on the Keck~II telescope under photometric conditions.
The image scale was 0\farcs211~pix$^{-1}$.
As the goal was to detect faint globular clusters in the cD halo, the total
integration time was 8400$\,$s; the seeing in the final stacked image was
0\farcs57.  The data were reduced as detailed by Blakeslee \etal\ (1997).
The photometric calibration of the image, determined
from Landolt (1992) standard stars observed the same night, agreed to
within 0.01~mag with that of a short $R$-band photometric exposure
taken with COSMIC (Dressler 1993) on the Palomar~5\,m~telescope.~ 

Figure~\ref{img} shows the central $3^\prime$ of the deep 8400\,s $R$-band
Keck image.  The arc is located $26\farcs9\pm0\farcs2$ from the center of
the cD at a position angle PA$\,=\,$141$^\circ$, very close to the major
axis of the cD.  The elliptical cD isophotes at radii $r_{\rm maj} \geq
19\arcsec$ along the major axis are oriented at 
PA${\,=\,}144^\circ{\pm}1^\circ$.  
(At $r_{\rm maj}<10\arcsec$, the cD isophotes are at
PA$\,\approx\,$157$^\circ$ and are less elliptical.)  
The cD halo has a surface brightness $\mu_R{\,=\,}22.9$\,mag~arcsec$^{-2}$ 
at the position of the arc.

\begin{figure*}[t]\epsscale{0.6}
\plotone{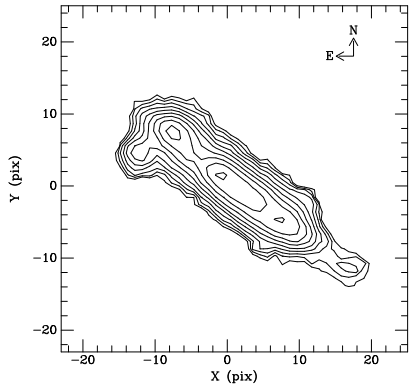}\caption{ \small
The isophotal contour map of the A2124 arc.
Contours are drawn in 0.4~mag increments with the brightest at 
$\mu_R=22.5$ mag arcsec$^{-2}$ and the faintest at 
$\mu_R=26.1$ mag arcsec$^{-2}$.  
A sky background of 20.69 mag arcsec$^{-2}$
has been subtracted, and
contributions from all other sources of light
have been removed by masking out the arc and interpolating
across it.  Because of the bright background, the precise
values of the faintest isophotes plotted are fairly uncertain.
The size of the map is $10\arcsec\times10\arcsec$.
\label{conts}}
\end{figure*}

The isophotal contour map plotted in Figure~\ref{conts}
reveals substructure in the A2124 arc.
The brightest intensity peak is located near the arc center, with
two fainter peaks near either end of the main part of the arc.
A faint tail extends $\sim\,$1\farcs5 further to the southwest
along the same line as the main part of the arc.
There is a stellar object located 1\farcs2 to the southeast of
the intensity peak at the northeast end of the arc; because it
does not follow the same locus, we believe this object is
unassociated with the arc (compare the inset in Figure~\ref{img}).
Its magnitude $m_R\sim24.5$ corresponds to the brightest globular
clusters in the~cD.  
After all sources of contamination are accounted for,
and corrected for 0.07 mag of Galactic extinction (\cite{SFD98}), 
the total $R$~magnitude of the arc is $m_R({\rm arc})=20.86\pm0.07$.

The axis ratio of the arc is poorly defined because of the substructure.
The axis ratio of the 23.4 mag arcsec$^{-2}$ isophote
(about half the intensity of the brightest isophote plotted) is
4.8, and would be $\sim\,$5.7 in the absence of seeing.
However, the total length of the arc including the faint 
tail to the southwest is at least 8\farcs5, and the full width
at half maximum normal to the arc at the central or northeast
intensity peaks is $\hbox{FWHM}=0\farcs83\pm0\farcs04$, which would be
$\hbox{FWHM}\sim0\farcs6$ in the absence of seeing.
Using these measures, the axis ratio is then $l/w\sim14$.
By measuring the deviation from lines drawn from one end of the arc to
the other, we estimate the radius of curvature to be 
$100\arcsec{\pm\,}20\arcsec$.

We also obtained $4{\times}400\,$s of $B$-band integration with COSMIC on
the Palomar 5\,m.  The seeing in the stacked $B$~image was 1\farcs7, making
it difficult to measure an accurate color.  For the brightest 7\arcsec\ of
the arc, we find $(B{-}R)_0 = 1.85\pm0.25$.

\subsection{Spectroscopic Data}

We obtained two 900\,s spectra of the arc with LRIS on
the night of 18~June 1998.  A 300 line grating blazed at 5000~\AA\
gave a dispersion of 2.44~\AA\,pix$^{-1}$ and wavelength coverage of
4000--9000 \AA.  The slit was 1\arcsec\ wide, yielding a resolution of
$\sim\,$10~\AA.  The slit was oriented at a PA of 54\fdg1, aligned with
the bright nucleus of the S0 galaxy 30\arcsec\ northeast of the
arc, as shown in Figure~\ref{img}.  The alignment
galaxy has an absorption spectrum that indicates it is 
an A2124 member at $z{\,=\,}0.0652$. 
Halogen flats and arc-lamp exposures were taken after
each spectrum, and standard IRAF routines were used for extraction and
wavelength calibration.  The spectra were flux calibrated using a
spectrum of Feige~110 taken at the end of the night.
Variable seeing and the presence of thin, patchy cirrus
makes the absolute calibration uncertain.~~ 

The raw two-dimensional spectra show one bright
emission line at the spatial position of the arc,
tilted obliquely with respect to the night sky lines.  The tilt 
is at least partly the result of the misalignment of the arc and the
slit.  On close inspection of the sky-subtracted
2-d spectra, two other much weaker emission lines are visible at
longer wavelengths with the same tilt as the brighter line.
The total tilt is consistent with the width of the slit, and the
length of the emission in the spatial direction is consistent with
the length of the main part of the arc in the $R$-band image.
This indicates that the arc was well-centered within the slit
and that the seeing was somewhat better than~1\arcsec.

\begin{figure*}[t]\epsscale{0.74}
\plotone{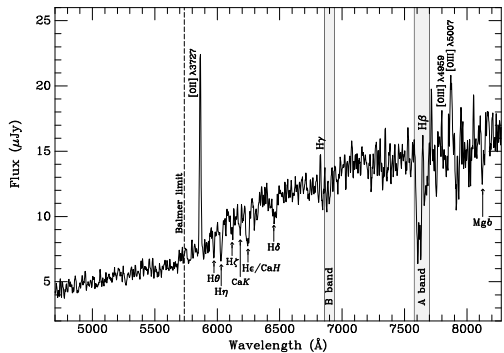}\caption{ \small
The sky-subtracted, calibrated 1-d spectrum of the arc, smoothed
with a 7.5\,\AA\ wide boxcar filter.
The shaded regions enclose the atmospheric $A$ and $B$
absorption bands.   The strong [OII]~$\lambda 3727$ line observed
at 5863\,\AA\ yields a redshift $z_{\rm arc} = 0.573$.
The [OIII]~$\lambda5007$ line is clearly detected, but is much
weaker than $\lambda 3727$ and somewhat obscured by residuals
from the numerous OH sky lines; [OIII]~$\lambda4959$ may also be present
at a low level.  H$\beta$ is redshifted into the 
$A$~band, but the putative emission has the same spatial signature 
as do the $\lambda3727$ and $\lambda5007$ lines, and thus is
apparently real.  A spike is visible at the redshifted position
of H$\gamma$, but the higher order Balmer lines appear in
absorption.   Mg$b$ absorption, indicating an older stellar
component, lands in a gap between sky emission lines and is visible.
The positions of the Ca~I $H\, \&\, K$ lines and the Balmer limit
are also shown.
\label{spec}}
\end{figure*}

Figure~\ref{spec} shows the final calibrated 1-d spectrum of the arc.
Identifying the strong emission line at 5863.5~\AA\ with
[OII]~$\lambda3727.5\,$\AA\ gives a redshift $z = 0.5730\pm0.0002$.
Identifying the line observed at 7798.2~\AA\ as [OII]~$\lambda5006.9$ gives
$z = 0.5723\pm0.0005$.  Unfortunately, H$\beta$ lands in the midst of the
atmospheric absorption of the Fraunhofer $A$~band.  There does appear to be
real emission (as evidenced by the 2-d spectrum) corresponding to H$\beta$
at $z = 0.5729$, but the irregular profile of the $A$~band makes the precise
position unreliable.  The other Balmer features identified in the figure
give an average $z\sim0.5726$.  Relying mainly on the [OII] line and
including calibration uncertainty, we adopt a redshift $z_{\rm arc} =
0.5728\pm0.0003$.  

We can identify some other absorption features in the arc spectrum.  Mg$\,b$
absorption is visible around $\sim\,$8135~\AA, in a gap between night sky
emission lines.  The Ca$\,K$ absorption line appears weaker than the Ca$\,H$
line because the latter is blended with H$\epsilon$.  Similarly, there is
not a strong 4000\,\AA\ break, but there is a change in spectral slope
corresponding to the Balmer limit near 5735\,\AA.

In comparison to the galaxy spectra in the atlases of Kennicutt (1992a)
and Liu \& Kennicutt (1995), the A2124 arc spectrum
most resembles those of the ``post-starburst'' galaxies
suggested to be the local analogs of distant E+A galaxies
(Dressler \& Gunn 1983). 
These are low ionization systems that may have strong [OII]
and H$\alpha$ (off the end of our spectrum) emission from ongoing
star formation, sometimes show weak emission at H$\beta$ and H$\gamma$,
but have the rest of the Balmer series appearing strongly in absorption;
they are thought to be evolved merger events.  This is what we see,
although the [OII] equivalent width measured for the arc
${\rm EW}_{\rm[OII]} = 30.1{\pm}1.6\,$\AA\ would place it 
among the youngest of these ``post-starburst'' galaxies.~ 

\section{Lens Models}

To model the cluster as a lens, we use an elliptical potential of the
form (Blandford \& Kochanek 1987)
$$
\psi(r^\prime) = 4 \pi {\left( {\sigma_{1D} \over c } \right)}^2
{{D_{LS}} \over {D_S}} \left[ {\left( 1 + { {r^\prime}^2 \over r_c^2 }
\right) }^{1 \over 2} - 1 \right] ,
$$
where $\sigma_{1D}$ is the line-of-sight velocity dispersion in the
limit $r^\prime \gg r_c$ for the spherical case, $D_S$ and $D_{LS}$ 
are angular size distances to the source and from 
the lens to the source, respectively, and $r_c$
defines a softening radius.  The ellipticity enters in the definition
of $r^\prime$,
$$
{r^\prime}^2 = (1 - \epsilon_p) x^2 + (1 + \epsilon_p) y^2 \,,
$$
where the coordinates $x,y$ are aligned along the major and minor axes
of the potential.  This represents a softened isothermal sphere for
$\epsilon_p = 0$, and tends to represent the shape of dark halo
potentials fairly well near the cores of clusters 
(e.g., Tyson \etal\ 1998). 
We chose the more limited relation with a fixed power
law, as there are few lensing constraints in this system to
differentiate various potential profiles.

We also chose to fix many of the model parameters using observational
constraints where available.  The orientation and ellipticity of the
cluster potential were set based on the measured $\theta_l,
\epsilon_l$ of cD halo light profile, using the approximation
$\epsilon_p = \epsilon_l/3$, following Mellier \etal\ (1993).  The
center of the cluster potential was fixed at the cD center.
Two additional spherical terms with $r_c = 0\farcs2$ were added 
to represent the central cusp of the cD itself and the halo of
the bright galaxy 12\arcsec\ northeast of the arc.  The cD
potential was normalized by setting the asymptotic dispersion
$\sigma_{1D} = 305$ \kms\ from the measured central dispersion (\cite{BT92}),
and  $\sigma_{1D} = 180$ \kms\ was assumed for the bright galaxy.

Our primary constraints for the lens model are the arc
radius, size, and orientation.  The lack of a counterarc suggests a
source position that lies outside the tangential and radial caustics
(e.g., Grossman \& Narayan 1988).
However, the length of the arc could imply
either a much longer arc opposite that is not observed, or a shorter
demagnified arc that would be ambiguous without redshift information.

\begin{figure*}[t]\epsscale{0.74}
\plotone{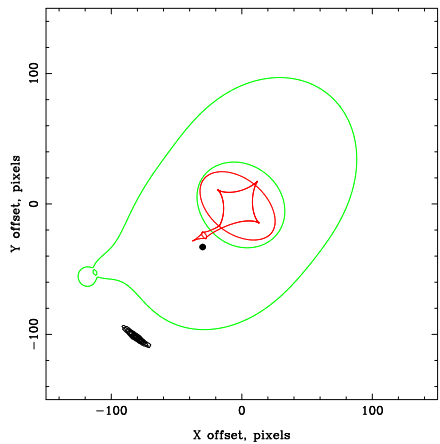}\caption{ \small
One of the best-fitting models of the Abell 2124 arc.  The cD lies at (0,0),
and the coordinates are in units of LRIS pixels (0\farcs211\,pix$^{-1}$) 
with north up and east to the left.  
The black contours indicate surface brightness of the
source and image; the caustics (critical surfaces in the source plane)
are shown in red, and the critical curves are shown in green.
}\end{figure*}

A family of models with a range of core radii and asymptotic
dispersions for the primary lens (cluster dark matter) were found that
reproduce the characteristics of the arc. 
These models are degenerate in $\sigma_{1D}$ and $r_c$
for determining the mass enclosed within 
the arc radius. 
Models with large core radii
tend to produce larger radial magnifications, leading to smaller intrinsic
source sizes and larger total magnifications.
Fixing the velocity dispersion to the Fadda
\etal\ (1996) value of 878 \kms, we find a best-fit model having 
$r_c=10$\arcsec\ (9 \hkpc) and a roughly round source 0\farcs4 in 
extent with an
unlensed position 6\farcs1 east and 6\farcs8 south of the cD
center. Figure~4 shows the source and image planes for this model,
along with the caustic and critical curve structure.  The
magnification of the source ranges between 9 and 13 for the family of
best-fit models.

\section{Discussion}

We find that quite reasonable values of the physical parameters of the
cluster and source can naturally produce a lensed arc at the observed radius.
Though our best-fit core radius for the cluster may seem small, it has been
shown that cluster potentials as measured by lensing in higher redshift
clusters tend to be steeper near the center than implied by models of
the X-ray emission (LeF\`evre \etal\ 1994; Waxman \& Miralda-Escude 1995).
The intrinsic dispersion of this cluster is also somewhat
ambiguous, and adopting dispersions nearer the larger estimate (\S1)
implies larger core radii.  The core radius is not strongly
affected by changing the model of the cD potential,
which is a relatively small perturbation on the cluster potential.
There are several faint sources in the
field that appear arc-like and may be strongly lensed; having
arcs at different radii would allow better constraints on the slope of the
potential near the core.

At $z{\,=\,}0.573$ and in the absence of lensing, 1\arcsec\ corresponds
to 4.3--3.7~\hkpc\ ($q_0{\,=\,}0$--0.5).
The observed 8\farcs5 arc length would then imply
a physical length of 53~kpc (for $q_0{\,=\,}0$, $h{\,=\,}0.7$),
or 64 kpc ($q_0{\,=\,}h{\,=\,}0.5$).  
With a $k$-correction of $-$0.4~mag, the observed magnitude 
$m_R = 20.86$ would correspond to $M_R \approx -21.3 + 5\log(h)$,
nearly $2L^*$ (Lin \etal\ 1996, with $R{-}r = -0.3$). 
The observed $(B{-}R)$ color is that of an E+A galaxy 
in which the merger/starburst
occurred $\lta\,$0.5~Gyr ago (Belloni \etal\ 1995;
Belloni \& R\"oser 1996).

We noted that the arc's spectrum resembles that of some
``post-starburst'' galaxies.  From the observed [OII] emission,
we can infer the ongoing star-formation rate (SFR). The total
[OII]~$\lambda3727$ line flux in the spectrum shown in Figure~\ref{spec}
is $(1.89{\pm}0.05)\times10^{-16}$ ergs~s$^{-1}$\,cm$^{-2}$.
We estimate the total flux to be larger by 15--20\% 
due to the placement of the arc in the slit and the better
seeing for the flux standard spectrum.  However, the presence
of thin cirrus makes the absolute
flux uncertain by at least another 10\%.  We therefore conservatively
adopt $f_{[{\rm OII}]} = (2.1{\pm}0.3)\times10^{-16}$ ergs~s$^{-1}$\,cm$^{-2}$.
This translates to a luminosity 
$L_{[{\rm OII}]} = (1.2{\pm}0.2)\times10^{41}$ $h^{-2}$\,ergs~s$^{-1}$.
Adopting the mean relation from Kennicutt (1998) and setting $h=0.7$
yields $\hbox{SFR} = 8.5\pm3.3\;\msun\,$yr$^{-1}$.
The SFR calibration based on [OII] equivalent width and observed
$B$~luminosity (Kennicutt 1992b, modified as in Kennicutt 1998)
is intrinsically much less certain but gives a similar result.

The $z{\,=\,}0.573$ galaxy would thus be an extremely large and 
luminous, linear or flattened galaxy with a lumpy morphology
and a substantial amount of ongoing star formation.  These points
would be indicative of a large merger event, except for the
narrow, slightly curved shape.  The extreme length if unlensed
(50--65\,kpc) by itself suggests that the arc is a gravitationally
magnified image of a smaller object with a more modest SFR.

Further observations might help to confirm the lens model and
constrain the cluster mass profile at small physical radii.  A weak
lensing analysis would help to determine the radial mass profile; such
an analysis is in progress but complicated by the extensive
globular cluster population.  A2124 would be an ideal target for
a wide-field imager on a 4\,m class telescope
(e.g., \cite{BRWL94}; Metzger, Luppino, \& Miyazaki 1995).
A search for a counterarc
via longslit spectroscopy might prove fruitful, since the
acceptable image positions would lie roughly along a line.  The
demagnification in models with additional images yields sources not
much fainter than $R=24$, and given the strong emission line in the
lensed galaxy, this could be detected with Keck spectroscopy.  In addition,
HST observations of the arc and possible additional arcs might show
substructure that would provide further lens mapping constraints.

\section{Summary and Conclusions}

We have identified an arc-like object at $z=0.573$ located 27\arcsec\
along the major axis from the center of the cD galaxy in A2124 ($z=0.066$).
The total magnitude of the arc is $m_R = 20.86\pm0.07$, corrected
for Galactic extinction, and the color is $(B{-}R)_0 = 1.85\pm0.25$.
There are three intensity peaks along its length 
and a faint extension $\sim$1\farcs5 to the southwest
that makes the total length 8\farcs5, and the axis ratio $\sim\,$14.
The spectrum shows strong [OII] emission with 
${\rm EW}_{\rm[OII]} = 30.1\pm1.6\,$\AA\ and strong higher-order Balmer
absorption; it thus resembles the spectra of some ``post-starburst'' 
galaxies.  If the object were unlensed,
the observed length would correspond to a physical size
of $\sim$50--65 kpc, and the observed [OII] emission would imply
$\hbox{SFR} = 8.5\pm3.3\;\msun\,$yr$^{-1}$ ($h{\,=\,}0.7$).
We believe it is more likely 
that the arc is the magnified image of a smaller background object,
and the true SFR is less than a tenth of this.
A simple lens model, appropriate to the core of A2124, produces
an arc radius equal to that observed.  This model and the
observed geometry strongly support the lensing hypothesis
for the A2124~arc; further observations may yield additional
constraints that could confirm or reject this hypothesis.
Finally, as the probability of such distant galaxies being lensed by
nearby clusters depends on the poorly-known shape of the inner cluster
potential, it would be worthwhile to try to find
these lenses in a systematic, rather than serendipitous, manner.

\acknowledgments

We thank Terry Stickel and Teresa Chelminiak for assistance 
at the telescope
and Judy Cohen, Bev Oke, and the rest of the team 
responsible for the Low Resolution Imaging Spectrograph.
We also thank C.-P.~Ma for valuable discussions.
Keck Observatory was made possible by the generous
financial support of the W.M. Keck Foundation.
J.P.B. is grateful to the Sherman Fairchild Foundation for support.
M.R.M's research was supported by Caltech.



\end{document}